\pdfoutput=1
\documentclass[sigconf]{acmart}

\usepackage{xspace}
\usepackage{enumitem}
\usepackage{caption}
\usepackage{subcaption}
\usepackage{multirow}
\usepackage{cleveref}
\usepackage{arydshln}
\usepackage{xspace}
\usepackage{url}
\usepackage{hyperref}
\usepackage[hyphenbreaks]{breakurl}

\AtBeginDocument{%
  \providecommand\BibTeX{{%
    \normalfont B\kern-0.5em{\scshape i\kern-0.25em b}\kern-0.8em\TeX}}}

\copyrightyear{2024}
\acmYear{2024}
\setcopyright{acmlicensed}\acmConference[ICSE-Companion '24]{2024
IEEE/ACM 46th International Conference on Software Engineering: Companion
Proceedings}{April 14--20, 2024}{Lisbon, Portugal}
\acmBooktitle{2024 IEEE/ACM 46th International Conference on Software
Engineering: Companion Proceedings (ICSE-Companion '24), April 14--20,
2024, Lisbon, Portugal}
\acmDOI{10.1145/3639478.3641226}
\acmISBN{979-8-4007-0502-1/24/04}

\begin{document}

\title{Program Decomposition and Translation with Static Analysis}

\author{Ali Reza Ibrahimzada}
\email{alirezai@illinois.edu}
\affiliation{
  \institution{\mbox{University of Illinois Urbana-Champaign}}
  \city{Champaign}
  \state{IL}
  \country{USA}
}


\begin{abstract}
  The rising popularity of Large Language Models (LLMs) has motivated exploring their use in code-related tasks. Code LLMs with more than millions of parameters are trained on a massive amount of code in different Programming Languages (PLs).
Such models are used for automating various Software Engineering (SE) tasks using prompt engineering. However, given the very large size of industry-scale project files, a major issue of these LLMs is their limited context window size, motivating the question of \textit{"Can these LLMs process very large files and can we effectively perform prompt engineering?"}.
Code translation aims to convert source code from one PL to another. In this work, we assess the effect of method-level program decomposition on context window of LLMs and investigate how this approach can enable translation of very large files which originally could not be done due to out-of-context issue. Our observations from $20$ well-known java projects and approximately $60K$ methods suggest that method-level program decomposition significantly improves the limited context window problem of LLMs by $99.5\%$. Furthermore, our empirical analysis indicate that with method-level decomposition, each input fragment on average only consumes $5\%$ of the context window, leaving more context space for prompt engineering and the output. Finally, we investigate the effectiveness of a Call Graph (CG) approach for translating very large files when doing method-level program decomposition.
\end{abstract}

\maketitle

\section{Introduction}

Machine learning has been widely used for automating various data-related tasks, and software engineering automation is no exception~\cite{copilot,cursor,palm,bard,chatgpt}. A subset of these machine learning tools called LLMs have shown significant improvements in several code related tasks~\cite{xia2022less,ibrahimzada2022perfect,xia2023conversational,pan2024lost,ibrahimzada2023automated,xia2023universal,dinella2022toga,rahman2023towards,chakraborty2023ranking,wei2023copiloting}. Prompt engineering constitutes a pivotal element contributing significantly to the achievements of recent LLMs~\cite{ye2023prompt}. However, context window, which contains both prompt to the LLM and the response to the prompt
from the LLM, is limited. So, prompt crafting, i.e., providing the minimum amount of information to maximize the gain from the LLM response, is crucial. This is not a certain issue for small
programs consisting of short methods. However, industry-scale software is typically large and complex, with a lot of dependencies between different components.

This work performs a large-scale study to investigate the effect of method-level program decomposition on the context window of LLMs. More specifically, our early experimental results from $20$ well-maintained Apache~\cite{apache} projects and roughly $60K$ methods demonstrate that real-life industry-scale software is very large and mostly cannot be processed by LLMs, demanding the necessity of fine-grained program decomposition techniques. Moreover, we show that when programs without decomposition fit in the context window of LLMs, they consume the majority of the context, leaving very little space for prompt engineering and the output. In contrast, method-level program decomposition improves the out-of-context issue of LLMs by $99.5\%$ with only consuming $5\%$ of the context window which ultimately enables processing of very large input files. Lastly, we perform a qualitative study on Apache Commons CLI~\cite{commons-cli} project when translating it using a CG approach to investigate the effectiveness of method-level program decomposition.
\section{Background}

Large Language Models (LLMs) have been extensively used in the domain of Natural Language Processing (NLP) and Programming Language Processing (PLP), achieving state-of-the-art performance in different tasks such as classification~\cite{yang2019xlnet}, translation~\cite{sutskever2014sequence}, automated program repair~\cite{xia2022less}, summarization~\cite{liu2019fine}. Prompt engineering involves providing the LLMs with minimal context to enhance their performance on any target task~\cite{ye2023prompt}. Recent studies have shown prompt crafting can significantly improve the quality of LLM responses~\cite{pan2024lost,xia2023conversational,wang2023leti}. Given the limited context window of state-of-the-art LLMs, effective decomposition of programs become an essential step when processing very large input files. Existing techniques on program decomposition involve program slicing~\cite{weiser1984slicing}. Starting from a subset of a program's behavior, slicing reduces that program to a minimal form which still produces that behavior. Each program fragment after slicing is independent and it should guarantee to represent the behavior of the original program. Another class of slicing techniques leverage dependency graph for decomposing very large programs~\cite{horwitz1988interprocedural}. These techniques mostly use flow graphs (CFG, DFG) for decomposing programs into smaller fragments.
\section{Approach}

In this section, we will discuss our technique for better program decomposition and translation using static analysis. This work performs a large-scale study to answer three main research questions (RQs). \textbf{First}, we were interested to understand if industry-scale real-life projects can fit in the context window of widely used LLMs. To answer this question, we downloaded $20$ well-maintained projects from Apache~\cite{apache}, located each stand-alone \texttt{\small .java} file from these projects and extracted their content. Next, we used the tokenizer of a widely-used open-source LLM called StarCoder~\cite{li2023starcoder} to tokenize the content of each file, and measure the number of files which do not fit in a $2K$ context window. We decided to compare all inputs against a $2K$ context size because most recent state-of-the-art LLMs come with a window size of $2048$ tokens~\cite{nijkamp2022codegen,fried2022incoder,li2023starcoder,allal2023santacoder}. \textbf{Second}, we wanted to investigate if method-level program decomposition can help with out-of-context problem of LLMs, and if effective prompt engineering is possible when encoding the decomposed method fragments. To address this concern, we performed static analysis on downloaded projects using CodeQL~\cite{codeql} to decompose each class in the projects into method fragments. As mentioned above, we used the same StarCoder tokenizer for tokenizing method fragments and measured the average number of tokens per method and the number of method fragments which do not fit in a $2K$ context window. For effective prompt engineering, we measured the amount of context size each method fragment on average would consume after decomposition. That is, the less space consumed by the input would enhance the performance of the model, leaving more context size for prompt engineering and output tokens. \textbf{Third}, we wanted to see if a program decomposition technique can help a simple translation approach in translating project-level programs. We used CodeQL for extracting the CG of each file and created a call dependency graph. Next, we implemented a simple translation technique which leverages the CG and translates the methods in a bottom-up manner. This approach makes sure to translate pure and independent method fragments first, and more dependent ones later by providing contextual information about independent methods.
\section{Evaluation}

\begin{table}
    \setlength{\tabcolsep}{4pt}
    \footnotesize
    \centering
    \caption{The effect of method-level program decomposition on a 2K context window model. The analysis has been done on 20 well-known Apache Commons projects.}
    \begin{tabular}{cccccc}
\hline
\textbf{Project} & \textbf{\begin{tabular}[c]{@{}c@{}}\% Files \\ \textgreater{}2K Tokens\end{tabular}} & \textbf{\# Methods} & \textbf{\begin{tabular}[c]{@{}c@{}}Avg. Tokens\\  / Method\end{tabular}} & \textbf{\begin{tabular}[c]{@{}c@{}}\% Methods \\ \textgreater{}2K Tokens\end{tabular}} & \textbf{\begin{tabular}[c]{@{}c@{}}\% 2K\\ Context\end{tabular}} \\ \hline
bcel            &  11.29\%                                     &  4,094                   &  70.42                             &  0.15\%                                   &  3.44\%                             \\
beanutils        &  29.84\%                                     &  2,675                   &  107.09                             &  0.07\%                                      &  5.23\%                             \\ 
cli            &  30.77\%                                     &  582                   &  97.91                             &  0.17\%                                       &  4.78\%                             \\
codec            &  48.30\%                                     &  1,788                   & 189.29                              &  0.84\%                                       &  9.24\%                             \\
collections            &  19.34\%                                     & 6,354                    &  74.37                             &  0.02\%                                       &  3.63\%                             \\
csv            &  27.08\%                                     &  871                   &  102.53                             &  0.11\%                                       &  5.01\%                             \\
daemon            &  27.78\%                                     &  60                   & 108.63                              &  0.00\%                                       &  5.30\%                             \\
dbcp            &  38.52\%                                     &   3,622                  &  63.02                             &   0.03\%                                      &  3.08\%                             \\
dbutils            &  13.54\%                                     &   869                  & 61.44                              &  0.00\%                                       &  3.00\%                             \\
fileupload            &  16.67\%                                     &  401                   &   77.8                            &   0.00\%                                      &   3.80\%                            \\
geometry            &  39.13\%                                     &   6,615                  &    124.93                           &  0.03\%                                       &  6.10\%                             \\
imaging            &  14.78\%                                     &  2530                   & 143.71                              &  0.20\%                                       &  7.02\%                             \\
io            &  22.07\%                                     & 5,957                    &  77.94                             &  0.07\%                                       &  3.81\%                             \\
jexl            &  25.70\%                                     &  3,967                   &  109.37                             &  0.20\%                                       & 5.34\%                              \\
lang            &  40.34\%                                     &  9,134                   &  103.33                             & 0.12\%                                        &  5.05\%                             \\
net            &  23.83\%                                     & 2,023                    & 98.22                              &  0.15\%                                       &  4.80\%                             \\
pool            &  22.68\%                                     & 1,377                    & 94.13                              & 0.00\%                                        & 4.60\%                              \\
rng            &  36.60\%                                     &  3,245                   &  139.69                             &  0.52\%                                       &  6.82\%                             \\
text            &  28.32\%                                     & 2,712                    &  99.85                             &   0.04\%                                      &   4.88\%                            \\
validator            &  38.00\%                                     & 1,181                    &  147.42                             &  0.17\%  &  7.20\%                             \\ \hline
\textbf{Average}            &  \textbf{27.73\%}                                     &  \textbf{3002.85}                   &  \textbf{104.55}                             & \textbf{0.14\%}                                        & \textbf{5.11\%}                              \\ \hline
\end{tabular}
    \label{table:results}
    \vspace{-10pt}
\end{table}

This section presents the results for our RQs. Our first RQ aims at understanding if industry-scale software can fit in the context window of recent LLMs. Column 2 under Table~\ref{table:results} shows the results for this study. As shown in the table, we observe roughly $30\%$ (min=$11.29\%$ and max=$48.30\%$) of input files from real-life projects do not fit in a $2K$ context window model. That is, one-third of files not only fits in the context window, they also do not leave any space for prompt engineering and new output generation. These results indicate an important problem of encoding very large programs by LLMs. To address this issue, we propose method-level program decomposition and discuss our results in RQ2.

Columns 3-6 in Table~\ref{table:results} show the results of RQ2. Motivated by the modularity of large-scale software, i.e., methods tend to be shorter and perform a single function, we decompose each \texttt{\small .java} file in the projects to a set of method fragments in order to address the out-of-context problem of LLMs. On average, each Apache project contains roughly $3,000$ methods under source and test directories, with each method consisting of approximately $100$ tokens. Doing such decomposition, we make sure that each fragment is independent and contains a subset of original program behavior as promised in program slicing~\cite{weiser1984slicing}. As indicated in Table~\ref{table:results}, method-level program decomposition improves the out-of-context problem by nearly $99.5\%$ ($27.73\%$ down to $0.14\%$), enabling encoding and processing of very large input files, which otherwise cannot be done due to their huge sizes. Furthermore, our proposed decomposition technique not only addresses the out-of-context issue, it also creates fragments which on average only consume $5\%$ of a $2K$ context window model. This will improve the quality of generated responses as LLMs tend to work well for shorter and well-engineered prompts~\cite{white2023prompt}.

In our last RQ, we wanted to investigate if our proposed decomposition technique can be incorporated with an SE task, i.e., code translation, to enable translation of large input files. Table~\ref{table:qualitative} shows a qualitative analysis when translating the source files of Apache Commons CLI~\cite{commons-cli} with method-level decomposition using its CG. We used an open-source model, i.e, StarCoder~\cite{li2023starcoder} for doing the translation. As shown in the table, vanilla translation of each file without any decomposition technique results in out-of-context error of 8 files. In contrast, when using method-level decomposition, all source files can be translated without any problems, by only consuming $3\%$ of context size (a 12$x$ improvement compared to no decomposition). Moreover, it is important to note that, our work in this paper focuses on enabling translation of large files, rather than correctly translating them which is a challenging problem by itself. Therefore, we do not validate the correctness of translations.

\begin{table}
    \setlength{\tabcolsep}{4pt}
    \footnotesize
    \centering
    \caption{The effectiveness of method-level decomposition when translating Apache Commons CLI using its Call Graph.}
    \begin{tabular}{cccc}
\hline
\textbf{\begin{tabular}[c]{@{}c@{}}Decomposition\\ Technique\end{tabular}} & \textbf{\begin{tabular}[c]{@{}c@{}}\# Source\\ Files\end{tabular}} & \textbf{\begin{tabular}[c]{@{}c@{}}\# Out-of-Context\\ Inputs\end{tabular}} & \textbf{\begin{tabular}[c]{@{}c@{}}\% Context\\ Occupied\end{tabular}} \\ \hline
No Decomposition                                                           & 22                                                                 & 8                                                                          & 36\%                                                                   \\
Method Decomposition                                                       & 22                                                                 & 0                                                                          & 3\%                                                                    \\ \hline
\end{tabular}
    \label{table:qualitative}
    \vspace{-10pt}
\end{table}

\section{Conclusion}
Program decomposition of large programs is essential for better prompt engineering of LLMs. In this work, we show that doing method-level program decomposition and CG-based translation enables translating of large input files. As part of our future work, we will explore how our approach can be combined with other techniques such as slicing and dependency graph decomposition.


\bibliographystyle{ACM-Reference-Format}
\bibliography{references}

\end{document}